\documentclass[
10pt,
aps,
pra,
twocolumn,
preprintnumbers,
superscriptaddress,
floatfix,
longbibliography,
nofootinbib
]{revtex4-2}
\usepackage{graphicx}
\usepackage{dcolumn}
\usepackage{bm}
\usepackage{amsmath}
\usepackage{amssymb}
\usepackage{txfonts}
\usepackage{hyperref}
\usepackage{color} 
\usepackage{xcolor}
\usepackage{listings}
\usepackage{cprotect}
\usepackage[ruled,lined]{algorithm2e}

\usepackage{tikz}
\usetikzlibrary{quantikz2}

\graphicspath{{./figures/}}

\newcommand{\Tr}{{\rm Tr}}

\newcommand{\e}{{\rm e}}

\newcommand{\im}{\mathrm{i}}
\newcommand{\diff}{\mathrm{d}}

\newcommand{\calR}{\mathcal{R}}
\newcommand{\calU}{\mathcal{U}}
\newcommand{\bE}{\mathbb{E}}

\hypersetup{
  breaklinks=true,
  colorlinks=true,
  linkcolor=[rgb]{0.60,0.00,0.00},
  citecolor=[rgb]{0.00,0.00,0.60},
  urlcolor=[rgb]{0.00,0.00,0.60},
  setpagesize=false
}

\definecolor{codegreen}{rgb}{0,0.6,0}
\definecolor{codegray}{rgb}{0.5,0.5,0.5}
\definecolor{codepurple}{rgb}{0.58,0,0.82}
\definecolor{backcolour}{rgb}{0.95,0.95,0.92}

\lstdefinestyle{mystyle}{
    backgroundcolor=\color{backcolour},   
    commentstyle=\color{codegreen},
    keywordstyle=\color{magenta},
    numberstyle=\tiny\color{codegray},
    stringstyle=\color{codepurple},
    basicstyle=\ttfamily\footnotesize,
    breakatwhitespace=false,         
    breaklines=true,                 
    captionpos=b,                    
    keepspaces=true,                 
    numbers=left,                    
    numbersep=5pt,                  
    showspaces=false,                
    showstringspaces=false,
    showtabs=false,                  
    tabsize=2
}

\begin{document}

\preprint{RIKEN-iTHEMS-Report-26}

\title{
Continuous-time evolution via probabilistic angle interpolation and its applications
}

\author{Tomoya Hayata}
\email{hayata@keio.jp}
\affiliation{Department of Physics, Keio University School of Medicine, 4-1-1 Hiyoshi, Kanagawa 223-8521, Japan}
\affiliation{RIKEN Center for Interdisciplinary Theoretical and Mathematical Sciences (iTHEMS), RIKEN, Wako 351-0198, Japan}
\affiliation{International Center for Elementary Particle Physics (ICEPP), The University of Tokyo, Hongo, 7-3-1 Hongo, Bunkyo-ku, Tokyo 113-0033, Japan}

\author{Yuta Kikuchi}
\email{yuta.kikuchi@quantinuum.com}
\affiliation{Quantinuum K.K., Otemachi Financial City Grand Cube 3F, 1-9-2 Otemachi, Chiyoda-ku, Tokyo, Japan}
\affiliation{RIKEN Center for Interdisciplinary Theoretical and Mathematical Sciences (iTHEMS), RIKEN, Wako 351-0198, Japan}

\begin{abstract}
We explore the applicability of a stochastic time-evolution algorithm based on probabilistic angle interpolation. To simplify the pre-processing of the algorithm, we take the continuous-time limit, thereby explicitly eliminating Trotter errors and streamlining the resource analysis. We also introduce a noise-mitigation method tailored to it. As demonstrations, we apply the algorithm to two representative problems: estimating the ground-state energy of the $H_3^+$ molecular Hamiltonian and computing out-of-time-ordered correlators in the sparse Sachdev--Ye--Kitaev model. We evaluate the protocol's performance through numerical simulations and experiments on a trapped-ion quantum computer, Quantinuum Reimei.
\end{abstract}

\date{\today}

\maketitle

\section{Introduction}
Current quantum hardware is subject to noise due to imperfect quantum operations and interactions with the environment.
Although quantum error correction is expected to lower the error rate, logical errors will remain non-negligible, particularly in the early stages of error-corrected quantum computing.
Consequently, the depth and width of coherently executable circuits remain limited.
This motivates the development of algorithms that reduce circuit size at the cost of other resources, such as sampling overhead and classical pre- and post-processing.
Stochastic quantum algorithms address this need by executing stochastically generated circuits that reproduce the target operation on average, and some of them offer a flexible trade-off between circuit size and sample complexity.

Hamiltonian simulation is a representative task that can benefit from stochastic implementations, and a variety of stochastic algorithms have been proposed~\cite{Campbell:2019fez,Childs2019,Chen2021,Wan:2021non,Nakaji:2023gze,Pocrnic:2023lrz,Granet2023,Watson:2024dvw,kiumi2025,Zeng2025simple}.
Among these, we focus on the stochastic time-evolution algorithm based on probabilistic angle interpolation (TE-PAI)~\cite{Koczor2023,kiumi2025} and adapt it to noisy quantum computers.
As detailed below, for a Hamiltonian $H$ written as a linear combination of Pauli strings, TE-PAI realizes channels $\rho\mapsto {\cal U}(t)[\rho]:=\e^{-\im t H}\rho\e^{\im t H}$ without time-discretization error after averaging over stochastically sampled unitary circuits.
We introduce a simplified classical pre-processing procedure by explicitly taking the continuous-time limit, together with a noise-mitigation method tailored to TE-PAI.

To demonstrate the feasibility of the algorithm, we estimate the ground-state energy of the $H_3^+$ molecular Hamiltonian using adiabatic ground-state preparation and the Hadamard test.
Hamiltonians in quantum chemistry are represented as a linear combination of many Pauli strings, and their $\ell_1$ norms are typically smaller than the number of terms.
Stochastic Hamiltonian simulation algorithms are therefore well-suited to this setting, and we numerically confirm these advantages.
In addition, we apply the method to compute out-of-time-ordered correlators (OTOCs) in the sparse Sachdev--Ye--Kitaev (SYK) model by classically simulating noisy quantum circuits.
Finally, we estimate the ground-state energy of $H_3^+$ on a trapped-ion quantum processor.

\section{Formalism}
\subsection{TE-PAI in the continuous-time limit}
\label{sec:continuous_tepai}
We start with an overview of TE-PAI~\cite{kiumi2025}. 
A quantum channel representing the Trotterized time evolution $\calU_\tau(t)$ with the Hamiltonian $H=\sum_{k=1}^{K} c_k P_k$ for Pauli strings $P_k$, time $t$, and the number of total Trotter steps $R$ is given by
\begin{align}
\label{eq:trotter_channel}
    \calU_\tau(t)[\rho]
    =
    \left(\prod_{k}^{\rightarrow} \e^{-\im\tau c_k P_k}\right)^R\rho  \left(\prod_{k}^{\leftarrow} \e^{\im\tau c_k P_k}\right)^R ,
\end{align}
with $\tau=\frac{t}{R}$, $\prod_{k}^{\rightarrow}A_k := A_1\dots A_K$ and $\prod_{k}^{\leftarrow}A_k := A_K\dots A_1$.
The key ingredient is the following identity for a Pauli-rotation channel $\calR_k(\theta_{k})[\rho]:=\e^{-\im \frac{\theta_k}{2}  P_k}\rho\e^{\im \frac{\theta_k}{2}  P_k}$~\cite{Koczor2023},
\begin{align}
    \calR_k(\theta_{k})
    =
    \gamma_1(|\theta_{k}|){\cal I} 
    + \gamma_2(|\theta_{k}|)\calR_k({\rm sgn}(\theta_{k})\Delta) 
    + \gamma_3(|\theta_{k}|)\calR_k(\pi) ,
\end{align}
with $\theta_k=2\tau c_k$, and the coefficients $\gamma_1(\theta) = \frac{\cos\frac{\theta}{2}\sin\frac{\Delta-\theta}{2}}{\sin \frac{\Delta}{2}}$, $\gamma_2(\theta) = \frac{\sin\theta}{\sin \Delta}$, $\gamma_3(\theta) = -\frac{\sin\frac{\theta}{2}\sin\frac{\Delta-\theta}{2}}{\cos \frac{\Delta}{2}}$.
Here, $\Delta\in(0,\pi)$ is a tunable parameter that controls the trade-off between the average gate count per circuit and sampling overhead as discussed below.
Regarding $p_l^{(k)}=|\gamma_l(|\theta_{k}|)|/\sum_i |\gamma_i(|\theta_{k}|)|$ as statistical weights, we stochastically sample a pair $(\gamma_{l(k,r)}, {\cal A}^r_k)$ from a set $\big\{(\gamma_1,{\cal I}),(\gamma_2,\calR_k({\rm sgn}(\theta_{k})\Delta)),(\gamma_3,\calR_k(\pi))\big\}$ for the $k$-th Hamiltonian term in the $r$-th Trotter step. It constructs an estimator of $\calU_\tau(t)$ as
\begin{align}
    \hat\calU_\tau(t)
    =
    \prod_{r=1}^R\prod_{k=1}^K \hat{\cal R}^r_k(\theta_{k}) ,
\end{align}
where 
\begin{align}
\label{eq:weight}
    \hat{\cal R}^r_k(\theta_{k})
    =
    \left(\sum_i |\gamma_i(|\theta_{k}|)|\right){\rm sgn}\left[\gamma_{l(k,r)}(|\theta_{k}|)\right]{\cal A}^r_k,
\end{align}
with $l(k,r)\in\{1,2,3\}$.
The Trotter circuit can be retrieved as the mean, $\bE[\hat\calU_\tau(t)]=\calU_\tau(t)$ with the normalization factor,
\begin{align}
\label{eq:overhead}
\begin{split}
    \sum_i |\gamma_i(|\theta_{k}|)|
    =
    \frac{\cos\left(\frac{\Delta}{2}-|\theta_k|\right)}{\cos\frac{\Delta}{2}}
    =
    1 + 2\tau|c_k| \tan \frac{\Delta}{2}
    +{\cal O}(\tau^2),
\end{split}
\end{align}
which contribute to sampling overhead.

As discussed in \cite{kiumi2025}, the step size $\tau$ can be made small to achieve an arbitrarily small time-discretization error without increasing the circuit size, which in turn introduces $R=t/\tau$ classical overhead to construct each stochastically sampled circuit.
Inspired by \cite{Granet2023}, we simplify the pre-processing by explicitly taking the continuous-time limit $\tau\to0$, which also allows us to eliminate the Trotter error and simulate the Hamiltonian evolution without bias. Introducing the rates of sampling $(\gamma_2,\calR_k({\rm sgn}(\theta_{k})\Delta))$ and $(\gamma_3,\calR_k(\pi))$, respectively,
\begin{align}
\label{eq:rates}
    r^{(k)}_2 
    = \lim_{\tau\to0}\frac{p_2^{(k)}}{\tau}
    = \frac{2|c_k|}{\sin \Delta},
    \qquad
    r^{(k)}_3 
    = \lim_{\tau\to0}\frac{p_3^{(k)}}{\tau}
    = |c_k|\tan \frac{\Delta}{2},
\end{align}
for each Hamiltonian term $P_k$, we propose the following protocol to estimate 
the expectation value of an operator $O$ after time evolution for time $t$:
\begin{enumerate}
\item Draw an integer $m$ according to a Poisson distribution with rate $r=\sum_{k=1}^K (r^{(k)}_2+r^{(k)}_3)$, i.e., with the probability $\e^{-rt}(rt)^m/m!$. Here, $m$ determines the number of Pauli-rotation channels applied in a sampled TE-PAI channel.
\item For each $i\in\{1,\dots,m\}$, draw a pair consisting of a Pauli-rotation channel and a sign from 
$\big\{
    \bigl(\calR_k(\operatorname{sgn}(\theta_k)\Delta),+1\bigr),
    \bigl(\calR_k(\pi),-1\bigr)
    \;|\;
    k=1,\dots,K
\big\}
$
with respective probabilities
$\big\{
    r_2^{(k)}/r,\,
    r_3^{(k)}/r
    \;|\;
    k=1,\dots,K
\big\}$. The products of the rotation channels and signs form a sampled TE-PAI channel $\hat{\cal U}(t)$ and a corresponding sign $s$.
\item We repeat the above two steps for $j\in\{1,\dots,N_{\rm s}\}$, with the sampled TE-PAI channel and sign obtained at the $j$-th step denoted by $\hat{\cal U}_j(t)$ and $s_j$. Using a quantum computer, we then measure the expectation value of $O$ after applying $\hat{\cal U}_j$ to an input state $\rho$. A weighted average of $N_{\rm s}$ expectation values gives an estimator of $\Tr[O\calU(t)[\rho]]$ as
\begin{align}
\label{eq:estimator}
    O_{\rm est}
    =
    \frac{\Gamma}{N_{\rm s}}\sum_{j=1}^{N_{\rm s}} s_j \Tr[O\hat{\cal U}_j[\rho]],
\end{align}
\end{enumerate}
where $\Gamma$ is the normalization factor given below by~\eqref{eq:normalization}.
See App.~\ref{app:derive_tepai} for a derivation.
Below, we call $\hat{U}(t)$ the unitary operator corresponding to a sampled TE-PAI channel $\hat{\cal U}(t)$, i.e., $\hat{\cal U}(t)[\cdot] = \hat{U}(t) [\cdot] \hat{U}^\dagger(t)$.

Since the normalization factor for each step is given by~\eqref{eq:overhead} before taking the continuous-time limit, the total normalization factor is
\begin{align}
\label{eq:normalization}
\begin{split}    
    \Gamma
    &=
    \lim_{\tau\to0}\prod_k\left(
        1 + 2\tau |c_k| \tan \frac{\Delta}{2}
        +{\cal O}(\tau^2)
    \right)^{t/\tau}
    \\
    &=
    \exp\left[2t\|c\|_1\tan \frac{\Delta}{2}\right],
\end{split}
\end{align}
with $\|c\|_1:=\sum_k |c_k|$, resulting in the sampling overhead $\Gamma^2$~\cite{Granet2023}.
Using the gate sampling rates~\eqref{eq:rates}, one finds the average number of Pauli-rotation gates for each circuit
\begin{align}
\label{eq:gate_counts}
    N_{\rm rot}
    :=
    t\|c\|_1\left(\frac{2}{\sin \Delta} + \tan \frac{\Delta}{2}\right)
    =
    t\|c\|_1\frac{3-\cos\Delta}{\sin \Delta}
    \approx
    \frac{2t\|c\|_1}{\Delta},
\end{align}
where the approximation holds for small $\Delta$.
While the sampling overhead exhibits the exponential growth in $t\|c\|_1$ for a fixed $\Delta$, it can be avoided by choosing $\Delta\sim\frac{1}{t\|c\|_1}$, which leads to $N_{\rm rot}={\cal O}(t^2\|c\|_1^2)$~\cite{Granet2023, kiumi2025}.
For comparison, we recall the required resources for other stochastic Hamiltonian simulation algorithms, qDRIFT~\cite{Campbell:2019fez} and TETRIS~\cite{Granet2025}.
The gate count of qDRIFT is ${\cal O}(t^2\|c\|_1^2/\delta)$ for a target bias $\delta$.
TETRIS uses $N_{\rm rot}={\cal O}(t^2\|c\|_1^2)$ gates per circuit without bias, which is the same as TE-PAI with the optimal $\Delta$.
These two methods as well as TE-PAI with the optimal $\Delta$ require ${\cal O}(1/\varepsilon^2)$ samples to achieve a target statistical uncertainty $\varepsilon$.
The crucial difference between TE-PAI and TETRIS is that the former approximates a time-evolution channel while the latter approximates a time-evolution unitary.
We employ both of them later in the application to ground-state energy estimation by leveraging their respective features.

\subsection{Error mitigation: zero-noise extrapolation}
We tailor the zero-noise extrapolation (ZNE)~\cite{Temme:2016vkz,Giurgica-Tiron:2020rcf} to the stochastic algorithm~\cite{Granet2025}.
Recall that the free parameter $\Delta$ does not modify the average channel $\mathbb{E}[\hat{\cal U}(t)]$ in the absence of hardware noise, yet has an impact on the gate counts in each circuit~\eqref{eq:gate_counts} and on the sampling overhead through $\Gamma$~\eqref{eq:normalization}.
In the protocol, we select two $\Delta$ values, $\Delta_1$ and $\Delta_2$.
Then we run the protocol with both angles and obtain the expectation values $y_1$ and $y_2$ from the corresponding circuits. 
We assume for simplicity that $\Delta$ is sufficiently small so that $N_{\rm rot}\approx 2t\|c\|_1/\Delta$.

We consider a linear extrapolation, assuming that the effect of noise on the expectation values is linear in the number of gates, $y_i=a+bN_{\rm rot}(\Delta_i)$, with fitting parameters $a,b$.
Under these assumptions, we obtain a mitigated expectation value by extrapolating $y$ to the $N_{\rm rot}\rightarrow0$ limit~\cite{Granet2025},
\begin{align}
\label{eq:ZNE}
    y_{\rm ZNE}=\frac{\Delta_1y_1-\Delta_2y_2}{\Delta_1-\Delta_2}.
\end{align}
We will discuss its utility and potential limitations in Sec.~\ref{sec:OTOC}.\footnote{
Similarly to the linear ZNE, one can consider other extrapolation strategies.
One of the representatives is the exponential extrapolation, where the assumption is that the effect of noise is of the form $y_i= a \e^{bN_{\rm rot}(\Delta_i)}$.
In that case, the exponential mitigation gives the estimator, ${\rm sgn}(y_1)\exp \left[\frac{\Delta_1\log |y_1|-\Delta_2 \log |y_2|}{\Delta_1-\Delta_2}\right]$, assuming $|y_1|>|y_2|$. While we focus on the linear ZNE in this work, the exponential ZNE can be more effective when the noise effect is larger due to the larger circuit size or the higher noise level of the quantum device.
}

\section{Applications}
\subsection{Application 1: ground-state energy estimation}
We estimate the ground state energy of the $H_3^+$ molecule, which is described by a 6-qubit Hamiltonian given by a linear combination of 42 Pauli terms (see \cite{Nutzel2025} and App.~\ref{app:adiabatic_prep} for the Hamiltonian).
\begin{figure}
\includegraphics[width=\linewidth]{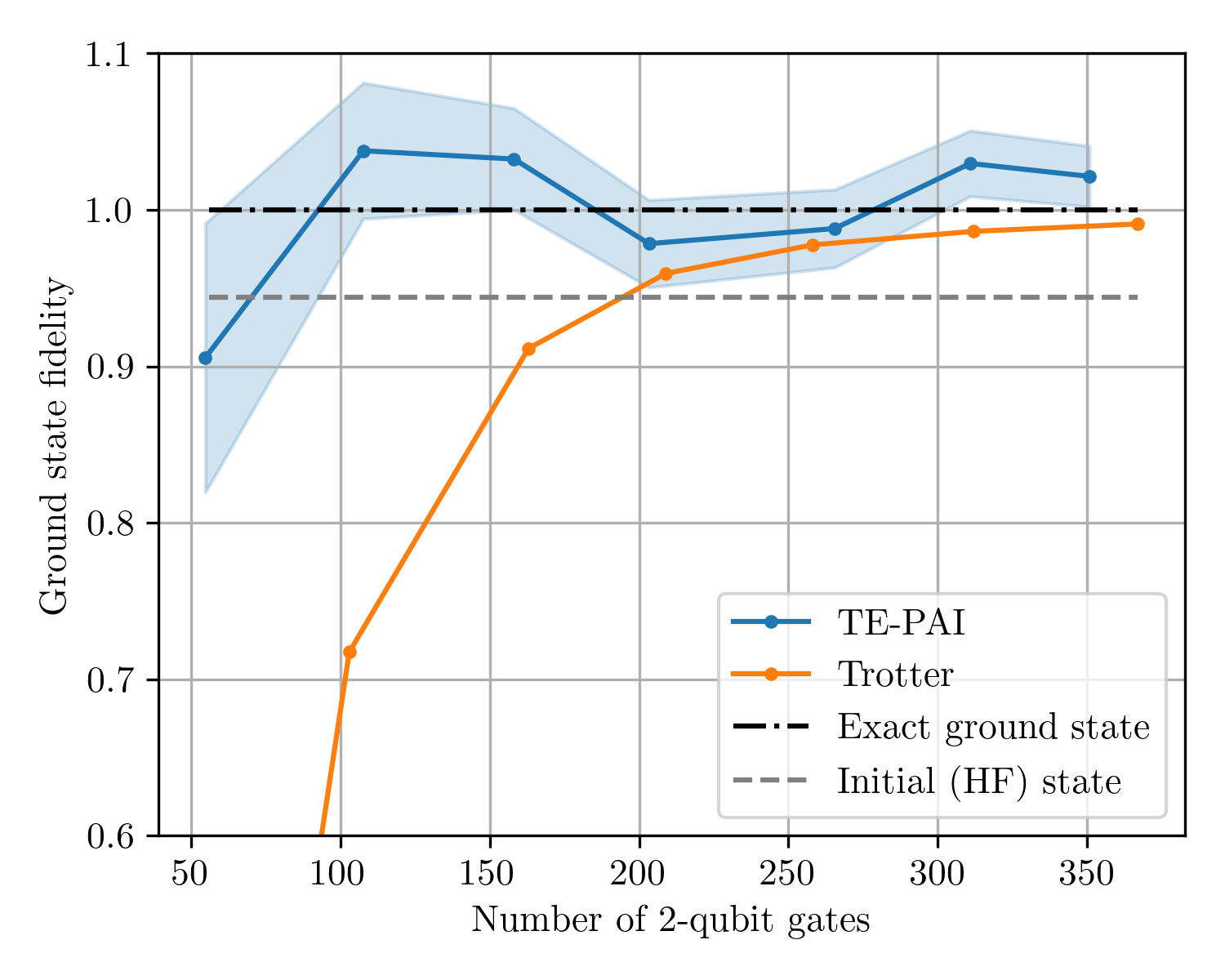}
\caption{\label{fig:trotter_vs_tepai_GSprep}
    Fidelities between the exact ground state and adiabatically prepared approximate ground states for the fixed adiabatic time $T=8$.
    The horizontal axis represents the (average) number of $R_{ZZ}(\theta)$ gates, which is the number of two-qubit gates in the circuit.
    Blue data represent the fidelity estimated with TE-PAI using 500 samples with statistical uncertainties represented by the opaque region. The data points correspond to $\Delta=\{\frac{1}{2},\frac{1}{4},\frac{1}{6},\frac{1}{8},\frac{1}{10},\frac{1}{12}, \frac{1}{14}\}$ with the average $R_{ZZ}$ gate counts $\{54.8,107.7,158.2,203.3,265.5,310.9, 350.8\}$.
    Orange data are computed with the statevector simulation of the corresponding Trotter circuit. The data points correspond to the number of Trotter steps $\{1,2,3,4,5,6,7\}$ with the $R_{ZZ}$ gate counts $\{56, 103, 163, 209, 258, 312, 367\}$.
}
\end{figure}
We first approximately prepare the ground state using the adiabatic evolution,
\begin{align}
    &U_{\rm Ad} := {\cal T}\exp\left[-\im T\int_0^1\diff s\, H_{\rm Ad}(s)\right],
    \\
    &H_{\rm Ad}(t/T) := H_{\rm HF} + \frac{t}{T}H_{\rm int},
\end{align}
for $t\in[0,T]$, starting with the unique ground state of the mean-field (Hartree-Fock) Hamiltonian $H_{\rm HF}$ denoted by $\ket{\rm HF}$. See Alg.~\ref{alg:tepai_adiabatic} in App.~\ref{app:adiabatic_prep} for the implementation details of adiabatic evolution with TE-PAI.
Figure~\ref{fig:trotter_vs_tepai_GSprep} compares TE-PAI and first-order Trotterization for the adiabatic state preparation, where the adiabatic time is fixed to $T=8$.
With TE-PAI, the fidelity is estimated by Eq.~\eqref{eq:estimator} with $O=\ket{\rm GS}\bra{\rm GS}$ for the exact ground state $\ket{\rm GS}$.
As seen in Fig.~\ref{fig:trotter_vs_tepai_GSprep}, the fidelity estimator can exceed unity due to statistical fluctuation because the normalization factor $\Gamma$ [Eq.~\eqref{eq:normalization}] is larger than 1.
To facilitate a resource comparison with quantum hardware experiments in mind, the circuits are decomposed with the gate set consisting of single-qubit and $R_{ZZ}(\theta)=\e^{-\im\frac{\theta}{2} ZZ}$ gates, and the number of two-qubit gates is given by the (average) number of $R_{ZZ}(\theta)$ gates.
Note that the resultant gate count is the number of rotation gates [Eq.~\eqref{eq:gate_counts}] weighted by the overhead of decomposing Pauli-rotation gates into the specified gate set. Since $R_{ZZ}(\theta)$ is the native two-qubit gate on the device we use and it is typically the major source of errors, the corresponding count is a more relevant metric to assess the implementation cost than the rotation-gate count [Eq.~\eqref{eq:gate_counts}].
The gate counts of TE-PAI and Trotterization are arranged by tuning $\Delta$ and the number of Trotter steps, respectively.
TE-PAI achieves the unit fidelity at the cost of sampling error represented by the shaded region, which shrinks as the gate count increases.
Its statistical error is smaller than the Trotter error until the 2-qubit gate count reaches $\sim250$.

Having prepared an approximate ground state, we apply the Hadamard test to estimate the ground state energy.
The whole circuit takes the form
\begin{align}
\label{circ:hadamard}
\begin{quantikz}
    \lstick{\ket{+}}
    &
    & \ctrl{1}
    & \meter{X/Y}
    \\
    \lstick{\ket{\rm HF}}
    & \gate{U_{\rm Ad}}
    & \gate{V_H}
    &
\end{quantikz}
\end{align}
with $V_H:=\e^{\im s H}$, where the measurements are performed in both $X$- and $Y$-bases. 
We employ TE-PAI to implement $U_{\rm Ad}$ as discussed above, and adopt another randomized algorithm, time evolution through
random independent sampling (TETRIS)~\cite{Granet2023}, to stochastically implement $V_H$.
Similarly to TE-PAI, TETRIS samples a sequence of unitary gates that approximates the target unitary operator $V_H$ (not a unitary channel) on average without bias, i.e., $\Gamma_{\rm TET}\bE[\hat{V}_H]=V_H$ with a known normalization factor $\Gamma_{\rm TET}$, making it suitable for the Hadamard test. See App.~\ref{app:tetris} for implementation details.
In practice, we execute $N_{\rm s}$ randomly sampled circuits, each of which is created by sampling a TE-PAI circuit for $U_{\rm Ad}$ and a TETRIS circuit for $V_H$.

Averaging over the measurement outcomes with an appropriate normalization factor, we calculate the real and imaginary parts of $\bra{\rm HF}U_{\rm Ad}^\dag V_H U_{\rm Ad}\ket{\rm HF}$, which allows us to read off
\begin{align}
\label{eq:eta}
\begin{split}
    \eta^{\pm}
    :=&\,
    {\rm Im}\big[
        \e^{-\im s E_{\rm HF}^{\pm}}\bra{\rm HF}U_{\rm Ad}^\dag V_H U_{\rm Ad}\ket{\rm HF}
    \big]
    \\
    \approx&\,
    \sin[s (E_{\rm exact} - E_{\rm HF}^{\pm})],
\end{split}
\end{align}
with $E_{\rm HF}^{\pm}:=E_{\rm HF}\pm\epsilon$ for a manually chosen parameter $\epsilon>0$ and $E_{\rm exact}$ is the exact ground state energy of $H$.
The approximate equality arises because $U_{\rm Ad}\ket{\rm HF}$ is an approximate ground state.
Thus, we can estimate the ground state energy $E_{\rm est}$ by~\cite{Granet2025practicality}
\begin{align}
\label{eq:E_est}
    E_{\rm est}
    =
    E_{\rm HF} - \frac{1}{s}\arctan\left(
        \tan(s\epsilon)\frac{\eta^++\eta^-}{\eta^+-\eta^-}
    \right).
\end{align}
The decay of the signal in~\eqref{eq:eta} due to noise cancels between the numerator and denominator in the argument of $\arctan(\cdot)$ in \eqref{eq:E_est}, making it robust against hardware noise.
\begin{figure}
\includegraphics[width=\linewidth]{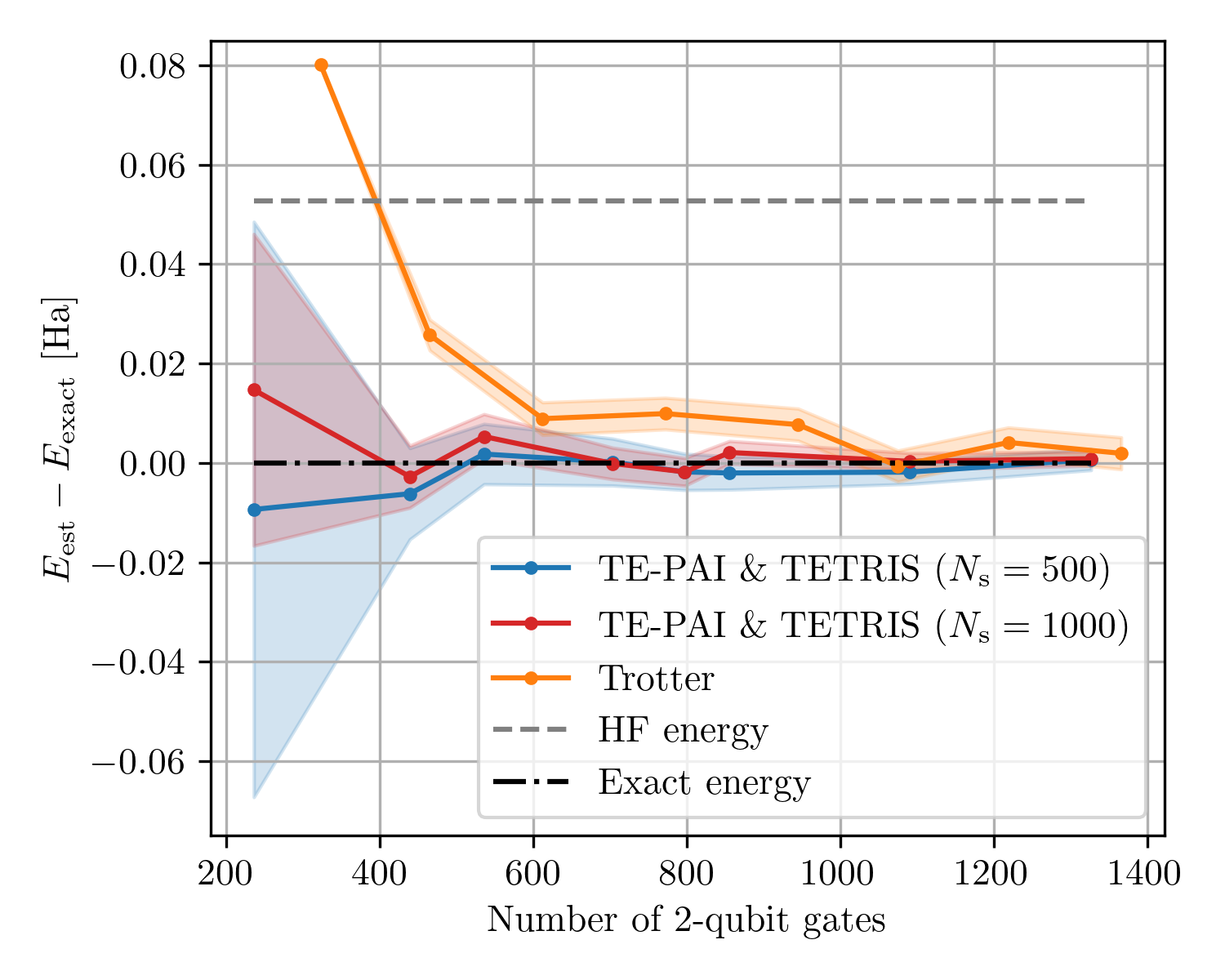}
\caption{\label{fig:trotter_vs_random_GSprep_Hadamard}
    Noiseless simulation of ground state energy estimation with TE-PAI \& TETRIS and first-order Trotterization.
    The adiabatic time is $T=8$, and the Hadamard test angle is $s=10$.
    Blue (red) data are computed using the former by sampling 500 (1000) circuits and executing 1 shot for each. The data points correspond to $\Delta=\{\frac{1}{2},\frac{1}{4},\frac{1}{5},\frac{1}{6},\frac{1}{7},\frac{1}{8},\frac{1}{10},\frac{1}{12}\}$ with the (average) $R_{ZZ}$ gate counts $\{235.8, 439.1, 535.7, 703.3, 797.3, 855.7, 1090.7, 1326.8\}$.
    Orange data are computed by executing 500 shots of a Trotter circuit. The data points correspond to the number of Trotter steps $\{1,2,3,4,5,6,7,8\}$ with the $R_{ZZ}$ gate counts $\{323, 465, 612, 772, 945, 1075, 1219, 1366\}$.
}
\end{figure}
Figure~\ref{fig:trotter_vs_random_GSprep_Hadamard} shows the noiseless calculation of the energy estimated by the stochastic algorithm with TE-PAI and TETRIS.
The adiabatic time is fixed to $T=8$ and the evolution time for the Hadamard test is $s=10$.
TE-PAI and TETRIS have a free angle parameter, both of which are set to $\Delta$.
For comparison, we also show the result obtained by using the first-order Trotterization for both adiabatic evolution and the Hadamard test. The time evolution operators for both the adiabatic evolution and the Hadamard test are Trotterized with the same number of Trotter steps.
In particular, the blue data (TETRIS \& TE-PAI) and orange data (Trotter) are computed with 500 shots/samples to facilitate an equal shot-budget comparison.
It shows that the stochastic algorithm estimates the energy without bias, albeit with sampling error, while the Trotterization exhibits negligible shot noise but introduces bias for small numbers of gates. Under the current setup with $N_{\rm s}=1000$, the former yields a slightly better estimate using up to approximately 1000 2-qubit gates.

\subsection{Application 2: OTOC of sparse SYK model}
\label{sec:OTOC}
Next, we numerically investigate the utility of ZNE by taking the sparse SYK model as an example, which is a quantum mechanical model with $N$ Majorana fermions $\psi_i$ ($\{\psi_i,\psi_j\}=2\delta_{ij}$) defined by~\cite{Xu2020,Garcia-Garcia2021,Caceres2021,Caceres2022,Caceres2023,Garcia-Garcia2023,Orman2024}
\begin{align}
\label{eq:sparseSYK}
    H = \sum_{i<j<k<l} p_{ijkl} J_{ijkl}\psi_{i}\psi_{j}\psi_{k}\psi_{l}.
\end{align}
The couplings $J_{ijkl}$ are independent random Gaussian variables with mean $0$ and variance $\mathrm{Var}[J_{ijkl}] = \frac{3!J^2}{N^{3}}$. $p_{ijkl}\in\{0,1\}$ are randomly sampled, being equal to $1$ with probability $p$, where $p$ controls the sparsity of the model. 
Motivated by an application to gravitational physics~\cite{Xu2020}, the probability is parametrized as $p=kN/\binom{N}{4}$ for an $N$-independent sparsity parameter $k$ and the variance is rescaled as $\mathrm{Var}[J_{ijkl}] = \frac{3!J^2}{pN^{3}}$. 
This leads to the average number of terms, $O(pN^4)=O(N)$, in the Hamiltonian~\eqref{eq:sparseSYK}.
The Hamiltonian on $N$ Majorana fermions can be expressed in terms of Pauli strings through the Jordan-Wigner transformation into $n=N/2$ qubits.
The resulting SYK Hamiltonians are thus all-to-all coupled, with long Pauli strings of average length $O(n)$.

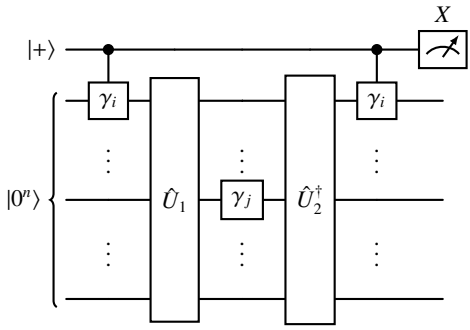
\begin{figure}
\begin{quantikz}[row sep=0.1cm, column sep=0.3cm]
    \lstick{$\ket{+}$}
    & \ctrl{1}
    & 
    &
    &
    & \ctrl{1}
    & \meter{X}
    \\
    \lstick[5]{$|0^n\rangle$}
    & \gate{\psi_i}
    & \gate[5]{\hat{U}_1}
    & \linethrough
    & \gate[5]{\hat{U}_2^\dag}
    & \gate{\psi_i}
    &
    \\
    & \vdots \wireoverride{n}
    & \wireoverride{n}
    & \vdots \wireoverride{n}
    & \wireoverride{n}
    & \vdots \wireoverride{n}
    & \wireoverride{n}
    \\
    &
    &
    & \gate{\psi_j}
    &
    &
    &
    \\
    & \vdots \wireoverride{n}
    & \wireoverride{n}
    & \vdots \wireoverride{n}
    & \wireoverride{n}
    & \vdots \wireoverride{n}
    & \wireoverride{n}
    \\
    &&&&&&
\end{quantikz}
\caption{\label{circ:OTOC}
The interferometric circuit to compute the real part of OTOC~\eqref{eq:def_OTOC}. The qubit at the top is the single-qubit ancillary register initialized to $\ket{+}$, while the remainder constitutes the system register with the input state $|0^n\rangle$. 
The unitary circuits $\hat{U}_1$ and $\hat{U}_2$ are sampled independently from TE-PAI circuits.
At the end of the circuit, the ancillary qubit is measured in the 
$X$ basis.
}
\end{figure}

OTOCs are used to probe the chaotic nature of quantum models~\cite{Larkin1969, Shenker2013, Roberts2014, Hosur2015, Swingle2018otoc, Xu2022}. Here, we calculate
\begin{align}
\label{eq:def_OTOC}
    {\rm OTOC}_{i,j} 
    := 
    \langle0^n|\psi_j(t)\psi_i\psi_j(t)\psi_i|0^n\rangle.
\end{align}
To this end, we employ the interference circuit given in Fig.~\ref{circ:OTOC}~\cite{Swingle2016, Swingle2018}.
For independently sampled TE-PAI circuit instances $\hat{U}_1$ and $\hat{U}_2$, the $X$-measurement calculates
\begin{align}
    {\rm Re}\big[\langle0^n|\hat{U}_1^\dag\psi_j \hat{U}_2\psi_i \hat{U}_2^\dag\psi_j \hat{U}_1\psi_i|0^n\rangle\big],
\end{align}
which, upon taking the weighted average of $\hat{U}_1$ and $\hat{U}_2$ according to \eqref{eq:weight}, recovers $\mathrm{Re}[{\rm OTOC}_{i,j}]$.
We remark that the sampling of the coupling parameters $J_{ijkl}$ and $p_{ijkl}$ and circuit sampling for TE-PAI are simultaneously performed, which saves the number of circuit executions compared to the naive approach of first sampling the Hamiltonian and then sampling TE-PAI circuits for each Hamiltonian instance.

\begin{figure}
\includegraphics[width=\linewidth]{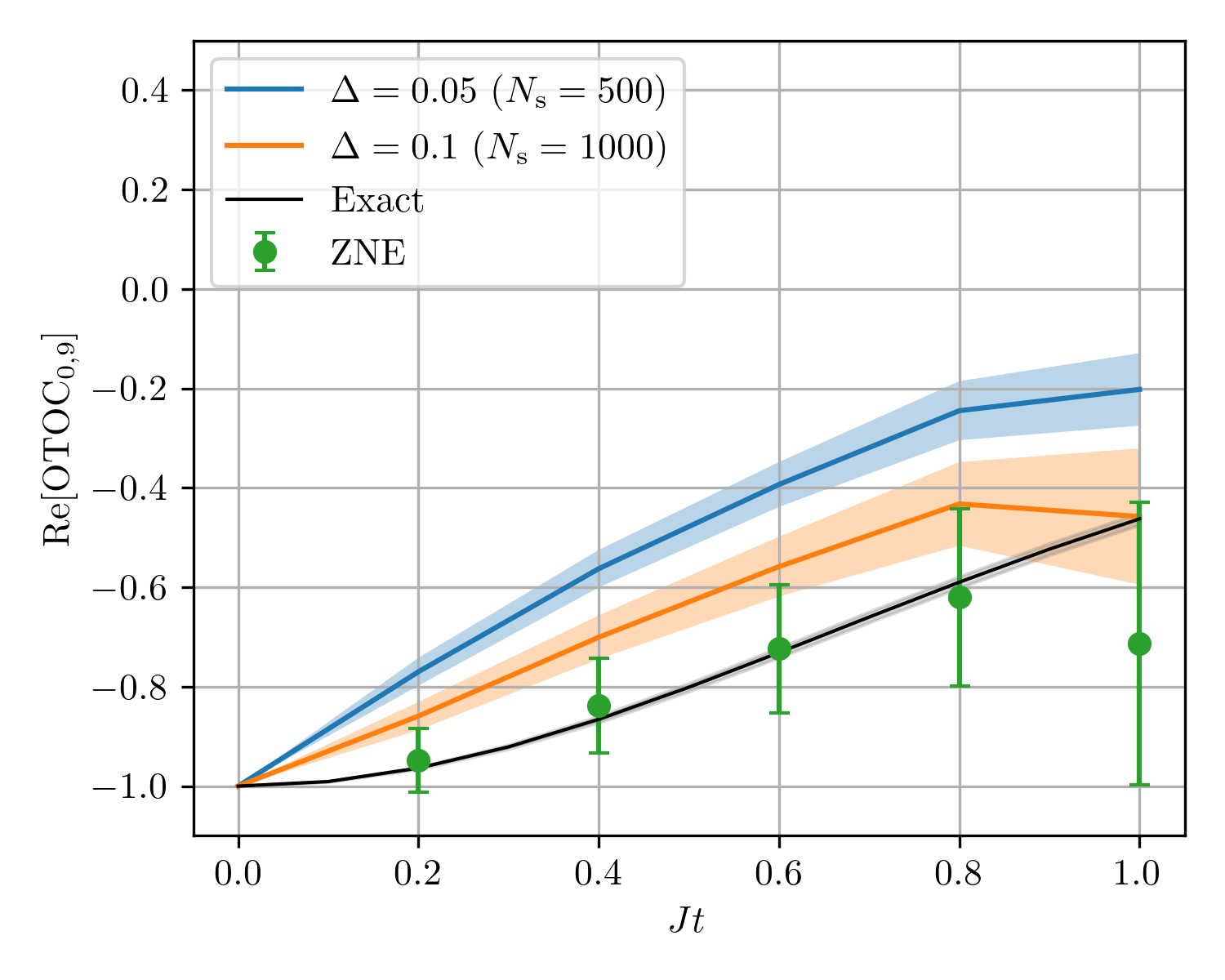}
\caption{\label{fig:SYK_ZNE_noisy_simulation}
    Noisy simulation of the interferometric circuit for OTOC calculations with ZNE (Fig.~\ref{circ:OTOC}). All $CX$s and single-qubit gates are subject to depolarizing noise of the parameters $10^{-3}$ and $10^{-5}$, respectively. The blue data are obtained by executing 500 circuits with $\Delta=0.05$, while the orange data are obtained by running 1000 circuits with $\Delta=0.1$. Each circuit is executed with 5 shots. The green data points represent the extrapolated values. The OTOC computed by noiseless simulation of continuous dynamics is shown by the black curve, which is obtained by averaging over 500 random realizations of $J_{ijkl}$ and $p_{ijkl}$, with the gray region representing the associated statistical error.
}
\end{figure}

We conduct a noisy simulation of the circuit. Here we decompose the circuits into the gate set consisting of single-qubit and $CX$ gates, and apply the depolarizing noise after each single-qubit gate with the strength $\lambda=10^{-5}$ and each $CX$ gate with $\lambda=10^{-3}$.
The SYK parameters are set to $N=18$ ($n=9$), and $k=2.3$ to compute the real part of the OTOC with $i=0$ and $j=9$.
For ZNE, we choose the TE-PAI angle parameter, $\Delta = 0.05, 0.1$, for all evolution times $Jt$.
To mitigate the noise effect, we measure the bottom register in Fig.~\ref{circ:OTOC} in the computational basis and post-select only when the parity of the bit string is odd. This post-selection discards the shots that violate the fermion parity symmetry of the Hamiltonian due to hardware noise.
Figure~\ref{fig:SYK_ZNE_noisy_simulation} shows the noisy simulation of the interferometric circuit for OTOC calculation with ZNE.
The simulation with a larger $\Delta$ suffers from less hardware noise but more sampling error. For each time $t$, we mitigate the error by ZNE.
While the proposed ZNE successfully restores the noiseless values within the statistical error, there exist various sources of error that are not captured by the present depolarizing noise model, which would limit the utility of the extrapolation~\eqref{eq:ZNE}.
To understand the impact of realistic hardware noise, additional numerical results are reported in App.~\ref{app:syk} along with the experimental results obtained using Reimei.

\subsection{Hardware experiment}
We experimentally demonstrate the ground-state energy estimation of the $H_3^+$ molecular Hamiltonian on the Quantinuum Reimei quantum computer, which operates on 20 all-to-all connected $^{171}$Yb$+$ qubits.

\begin{figure}
\includegraphics[width=\linewidth]{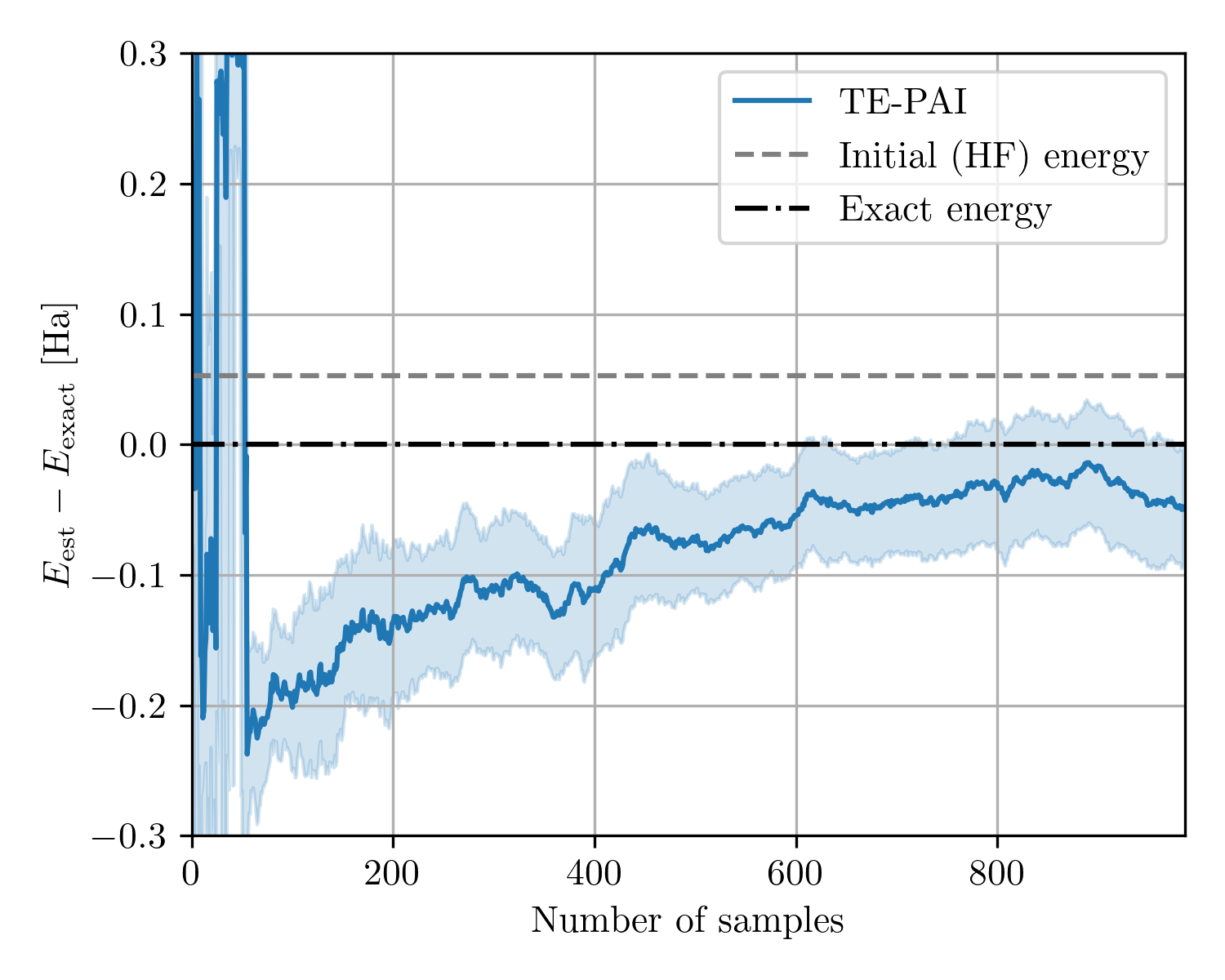}
\caption{\label{fig:energy_estimation_reimei}
    Energy estimate relative to the exact ground state energy calculated with Reimei against the number of samples. The blue curve shows the mean of $E_{\rm est}-E_{\rm exact}$ with the shaded region representing the statistical error.
    The dashed line corresponds to $E_{\rm HF}-E_{\rm exact}$, which is the energy of the initial state $\ket{\rm HF}$ relative to the exact ground state energy.
}
\end{figure}

We first evolve $\ket{\rm HF}$ under the adiabatic Hamiltonian $H_{\rm Ad}(s)$ with $T=8$ using TE-PAI, and then apply the Hadamard test with $s=10$ to estimate the ground state energy using TETRIS. TE-PAI and TETRIS have a free angle parameter, both of which are set to $\Delta$.
We execute $N_{\rm s}=1000$ randomly sampled circuits, consisting of 500 instances with $\Delta=0.1$ and 500 instances with $\Delta=0.2$, respectively, each of which is created by sampling a TE-PAI unitary for $U_{\rm Ad}$ and a TETRIS unitary for $V_H$.
We measure the bottom register in the circuit~\eqref{circ:hadamard} in the computational basis and post-select only when the parity of the bit string is even.
Figure~\ref{fig:energy_estimation_reimei} shows the energy estimate relative to the exact ground state energy against the number of samples.
The estimate approaches the exact value as the number of samples increases. However, the statistical uncertainty, which stems from the vanishing signal due to hardware noise, makes it inconclusive whether the estimate is more accurate than that of the initial state $\ket{\rm HF}$, which has a relative energy of $\sim0.05\,{\rm Ha}$.
We also observe bias in the energy estimation, in particular for small numbers of samples, despite the unbiased nature of the time-evolution algorithm. This is attributed to the non-linearity of the energy estimator in Eq.~\eqref{eq:E_est}, which is pronounced by the suppression of the signal $\eta^\pm$ by hardware noise because it makes the argument of the arctan function large due to small $|\eta^+-\eta^-|$.

\section{Discussion}
We have formulated TE-PAI directly in the continuous-time limit, thereby eliminating Trotter bias at the channel level while simplifying classical pre-processing.
The resulting picture makes the resource trade-off transparent: the angle parameter $\Delta$ controls circuit depth through Eq.~\eqref{eq:gate_counts} and sampling overhead through Eq.~\eqref{eq:normalization}.
This tunability is useful in the noisy regime, where one often prefers shallower circuits even with additional sampling, and it naturally enables a stochastic-ZNE strategy by comparing runs at different $\Delta$ values.

Our two applications showcase the approach's strengths.
For the $H_3^+$ simulation, TE-PAI-based adiabatic preparation combined with TETRIS in the Hadamard test yields an unbiased energy estimate and reaches accurate values with moderate two-qubit gate counts.
For sparse SYK OTOCs, the same stochastic framework supports interferometric estimation in the presence of long, nonlocal Pauli strings, and noisy simulation shows that extrapolation across $\Delta$ values can partially recover the noiseless trend.
The hardware demonstration on Reimei exhibits a similar trend, but the vanishing signal due to hardware noise makes it challenging to suppress the statistical error.

The present experimental demonstration of the protocol remains limited by sampling cost.
Although choosing a large $\Delta$ suppresses hardware noise by reducing per-circuit depth, it increases the number of required samples and can lead to large fluctuations, while a smaller $\Delta$ has the opposite behavior.
Improving this balance is the main practical challenge.
Promising directions include adaptive scheduling of $\Delta$ over evolution time, variance-aware allocation of shots between extrapolation points, and the use of variance reduction methods such as the echo verification~\cite{OBrien2021, Huo2021, Cai2021, Polla2023}.
Extending the benchmark set to larger molecules and larger sparse-SYK instances, and comparing against alternative randomized simulation schemes under equal hardware budgets, will clarify where continuous-time TE-PAI offers the largest advantage.


\begin{acknowledgments}
We thank Kazuhiro Seki for collaboration on the early stages of this work. We also thank Etienne Granet and Connor Lenihan for their feedback on the manuscript.
This work was based on results obtained from a project, JPNP20017, commissioned by the New Energy and Industrial Technology Development Organization (NEDO).
We acknowledge support from the RIKEN TRIP initiative (RIKEN Quantum).
T.~H.~was supported by JSPS KAKENHI Grant No.~JP24K00630, JP25K01002, and JST COI-NEXT (Grant No. JPMJPF2221).
\end{acknowledgments}

\bibliographystyle{apsrev4-2}
\bibliography{tepai}

\clearpage
\onecolumngrid
\appendix

\section{Derivation of TE-PAI in continuous time limit}
\label{app:derive_tepai}

We show that TE-PAI implements the unitary channel in Eq.~\eqref{eq:trotter_channel} in expectation in the $\tau\to0$ limit.
The time-evolution channel under a Hamiltonian $H=\sum_{k=1}^{K}c_kP_k$ for time $t$ can be converted as follows,
\begin{align}
    \calU(t)[\rho]=\e^{-\im tH}\rho\e^{\im tH}
    &=
    \lim_{R\to\infty}
    \left(\prod_{k=1}^K \calR_k(\theta_{k})\right)^{R}[\rho]
    \nonumber\\
    &=
    \lim_{R\to\infty}
    \left(\prod_{k=1}^K \Gamma_k\left(
        (1-p^{(k)}_2-p^{(k)}_3){\cal I} 
        + p^{(k)}_2\,\calR_k({\rm sgn}(\theta_{k})\Delta) 
        - p^{(k)}_3\,\calR_k(\pi)
    \right)\right)^{R}[\rho]
    \nonumber\\
    &=
    \Gamma
    \lim_{R\to\infty}
    \left(1-\tau\sum_{k=1}^K 
        \left(
            (r^{(k)}_2+r^{(k)}_3){\cal I} 
            - r^{(k)}_2\calR_k({\rm sgn}(\theta_{k})\Delta) 
            + r^{(k)}_3\calR_k(\pi)
        \right)
        + {\cal O}(\tau^2)
    \right)^{R}[\rho]
    \nonumber\\
    &=
    \Gamma
    \e^{-\sum_k(r^{(k)}_2+r^{(k)}_3)t}
    \exp\left(t\sum_{k=1}^K \left(
        r^{(k)}_2\calR_k({\rm sgn}(\theta_{k})\Delta) 
        - r^{(k)}_3\calR_k(\pi)
    \right)\right)
    [\rho]
\label{eq:derive_tepai_1}
\end{align}
with $\theta_k=2\tau c_k$, $\tau=t/R$, $\Gamma_k=\sum_i|\gamma_i(|\theta_k|)|$, and $r_{l}^{(k)}=\lim_{\tau\to0}p_{l}^{(k)}/\tau=\lim_{\tau\to0}|\gamma_{l}(|\theta_k|)|/(\Gamma_k \tau)$. The normalization factor $\Gamma$ is given by Eq.~\eqref{eq:normalization} in the main text.
Defining the total rate $r=\sum_k(r^{(k)}_2+r^{(k)}_3)$ and expanding the exponential function in Eq.~\eqref{eq:derive_tepai_1}, we find
\begin{align}
\label{eq:derive_tepai_2}
    \calU(t)[\rho]
    &=
    \Gamma
    \e^{-rt}
    \sum_{m=0}^{\infty}\frac{(rt)^m}{m!}
    \left(\sum_{k=1}^K \left(
        \frac{r^{(k)}_2}{r}\calR_k({\rm sgn}(\theta_{k})\Delta) 
        - \frac{r^{(k)}_3}{r}\calR_k(\pi)
    \right)\right)^m
    [\rho].
\end{align}

The right-hand side of Eq.~\eqref{eq:derive_tepai_2} can be seen as a quasi-probability decomposition of the time-evolution channel, i.e.,
\begin{align}
    \calU(t)[\rho]
    &=
    \Gamma\sum_{(m,\boldsymbol{k},\boldsymbol{l})} q_{(m,\boldsymbol{k},\boldsymbol{l})} s_{(m,\boldsymbol{k},\boldsymbol{l})}\hat{\calU}_{(m,\boldsymbol{k},\boldsymbol{l})}[\rho]
    =:
    \Gamma\,\mathbb{E}_{{(m,\boldsymbol{k},\boldsymbol{l})}}[s_{(m,\boldsymbol{k},\boldsymbol{l})}\hat{\calU}_{(m,\boldsymbol{k},\boldsymbol{l})}[\rho]],
\end{align}
where $\boldsymbol{k}=(k_1,\ldots,k_m)\in\{1,\ldots,K\}^m$
specifies the sequence of Pauli indices, and
$\boldsymbol{l}=(l_1,\ldots,l_m)\in\{2,3\}^m$
specifies the channel type at each position in the sequence:
$l_j=2$ corresponds to
$\calR_{k_j}(\operatorname{sgn}(\theta_{k_j})\Delta)$,
whereas $l_j=3$ corresponds to $\calR_{k_j}(\pi)$.
The quantity $q_{(m,\boldsymbol{k},\boldsymbol{l})}$ is the probability of sampling a channel $\hat{\calU}_{(m,\boldsymbol{k},\boldsymbol{l})}$, which is a product $m$ Pauli-rotation channels, and $s_{(m,\boldsymbol{k},\boldsymbol{l})}$ is the sign multiplying the channel.
Therefore, by sampling $\{s_{(m,\boldsymbol{k},\boldsymbol{l})}, \hat{\calU}_{(m,\boldsymbol{k},\boldsymbol{l})}\}$ according to the probabilities $q_{(m,\boldsymbol{k},\boldsymbol{l})}$, one can simulate the continuous time-evolution to unbiasedly estimate the expectation value of any observable $O$ as 
\begin{align}
    \Tr[O \calU(t)[\rho]] 
    = 
    \Gamma\,\mathbb{E}_{(m,\boldsymbol{k},\boldsymbol{l})}[s_{(m,\boldsymbol{k},\boldsymbol{l})}\Tr[O\,\hat{\calU}_{(m,\boldsymbol{k},\boldsymbol{l})}[\rho]]].
\end{align}

To sample the sign and the channel according to the probability $q_{(m,\boldsymbol{k},\boldsymbol{l})}$, we follow the procedure below.
\begin{enumerate}
    \item Draw an integer $m$ (number of Pauli-rotation gates per circuit) according to a Poisson distribution with rate $r$, i.e., with the probability $\e^{-rt}(rt)^m/m!$.
    \item For each $i\in\{1,\dots,m\}$, independently draw one of the $2K$
 pairs of channel and sign
\[
\Big\{
    \big(\calR_k(\operatorname{sgn}(\theta_k)\Delta),+1\big),
    \big(\calR_k(\pi),-1\big)
    \;\big|\;
    k=1,\dots,K
\Big\}
\]
with respective probabilities
\[
\Bigg\{
    \frac{r_2^{(k)}}{r},
    \frac{r_3^{(k)}}{r}
    \;\bigg|\;
    k=1,\dots,K
\Bigg\},
\]
and apply the sampled channels in increasing order of $i$.
\end{enumerate}
Using the sampled TE-PAI channels and signs, one can estimate the expectation value of an observable $O$ after time evolution under a time-independent Hamiltonian $H$ and for time $t$ as prescribed in Sec.~\ref{sec:continuous_tepai} of the main text.

\section{Adiabatic ground state preparation of molecular Hamiltonian}
\label{app:adiabatic_prep}

Following the setup of~\cite{Nutzel2025}, we estimate the ground state energy of the $H_3^+$ molecule given by the 6-qubit Hamiltonian
\begin{align}
\label{eq:H3+_hamiltonian}
\begin{split}
    H = &- 2.77 IIIIII  \\
    &+ 5.81 \times 10^{-1} IIIIIZ + 5.81 \times 10^{-1} IIIIZI + 5.81 \times 10^{-1} IIIZII \\
    &+ 5.81 \times 10^{-1} IIZIII + 7.48 \times 10^{-1} IZIIII + 7.48 \times 10^{-1} ZIIIII \\
    &+ 2.46 \times 10^{-2} IIIIZZ - 2.01 \times 10^{-3} IIIZIZ - 1.87 \times 10^{-2} IIZIIZ \\
    &- 1.87 \times 10^{-2} IIIZZI + 2.46 \times 10^{-2} IIZZII - 2.81 \times 10^{-2} IYIYII \\
    &- 2.81 \times 10^{-2} IXIXII - 2.01 \times 10^{-3} IIZIZI + 7.96 \times 10^{-4} ZZIIII \\
    &+ 2.81 \times 10^{-2} YZYIIZ + 2.81 \times 10^{-2} XZXIIZ + 2.16 \times 10^{-2} IIYXXY \\
    &- 2.16 \times 10^{-2} IIYYXX - 2.16 \times 10^{-2} IIXXYY + 2.16 \times 10^{-2} IIXYYX \\
    &- 2.81 \times 10^{-2} YZYZII - 2.81 \times 10^{-2} XZXZII + 2.81 \times 10^{-2} IYXIXY \\
    &+ 2.81 \times 10^{-2} IXXIXX + 2.81 \times 10^{-2} IYYIYY + 2.81 \times 10^{-2} IXYIYX \\
    &+ 3.83 \times 10^{-2} YXIIXY + 2.81 \times 10^{-2} IYZYZI - 3.83 \times 10^{-2} YYIIXX \\
    &- 3.83 \times 10^{-2} XXIIYY + 2.81 \times 10^{-2} IXZXZI + 3.83 \times 10^{-2} XYIIYX \\
    &+ 3.83 \times 10^{-2} YXXYII - 3.83 \times 10^{-2} YYXXII - 3.83 \times 10^{-2} XXYYII \\
    &+ 3.83 \times 10^{-2} XYYXII - 2.81 \times 10^{-2} XIIYYX + 2.81 \times 10^{-2} XIIXYY \\
    &+ 2.81 \times 10^{-2} YIIYXX - 2.81 \times 10^{-2} YIIXXY,
\end{split}
\end{align}
which consists of 41 non-identity Pauli terms.
Since the $H_3^+$ Hamiltonian conserves the total electron number 2, we only consider the subsector with $\sum_iZ_i/2=1$, where $\ket{1}$ and $\ket{0}$ represent the occupied and unoccupied states of each spin-orbital.
The Hartree-Fock (HF) state $\ket{\rm HF}=\ket{110000}$ is the ground state of the Hamiltonian $H_{\rm HF}$ composed only of single-$Z$ terms in Eq.~\eqref{eq:H3+_hamiltonian}.

We prepare the ground state of $H$ by adiabatically evolving the HF state under the time-dependent Hamiltonian,
\begin{align}
    H_{\rm Ad}(t/T) := H_{\rm HF} + f(t/T)H_{\rm int},
\end{align}
where $H_{\rm int}:=H-H_{\rm HF}$ and $f(s)=s$.
We employ the continuous TE-PAI algorithm to implement,
\begin{align}
    U_{\rm Ad} := {\cal T}\exp\left[-\im T\int_0^1\diff s\, H_{\rm Ad}(s)\right].
\end{align}
For notational convenience, we write the Pauli-decomposition of the Hamiltonians as
\begin{align}
\label{eq:H_decomposition}
    H_{\rm HF} = \sum_{k=1}^{K_{\rm HF}} b_k Z_k,
    \qquad
    H_{\rm int} = \sum_{k=1}^{K_{\rm int}} a_k P_k,
\end{align}
and define $z(s):=\int_0^s\diff s' f(s')$ and $\|a\|_1=\sum_k |a_k|$.  We also define $z^{-1}(s)$ to be the inverse function of $z(s)$, i.e., $z(z^{-1}(s))=z^{-1}(z(s))=s$.
We modify the algorithm presented in the main text as shown in Alg.~\ref{alg:tepai_adiabatic}.
\begin{algorithm}
\caption{Continuous TE-PAI for adiabatic evolution}
\label{alg:tepai_adiabatic}
\SetKwInOut{Input}{input}
\SetKwInOut{Output}{output}
\SetKwFor{ForEach}{for each}{do}{end}
    \Input{
        Adiabatic time $T>0$,
        rotation angle $\Delta\in (0, \pi)$,
        number of TE-PAI samples $N_{\rm s}$,
        the Hamiltonian $H=H_{\rm HF}+H_{\rm int}$,
        the operator $O$ to be measured,
        and the input state $\ket{\rm HF}$.
    }

    \For{$j=1$ \KwTo $N_{\rm s}$}{
        $m\gets$ draw from Poisson distribution with parameter $\|a\|_1 z(1) T\big(\frac{2}{\sin\Delta} + \tan\frac{\Delta}{2}\big)$
        \tcp*{draw number of Pauli-rotation gates}
        $gates \gets$ empty set\;
        \For{$i=1$ \KwTo $m$}{
            $t\gets Tz^{-1}(u)$ with $u$ drawn uniformly from the interval $[0, z(1)]$\;
            $k\gets$ draw from $\{1,\dots,K_{\rm int}\}$ with probability $|a_{k}|/\|a\|_1$\;
            $l\gets$ draw from $\{2, 3\}$ with probabilities $\frac{2}{\sin\Delta}/\big(\frac{2}{\sin\Delta} + \tan\frac{\Delta}{2}\big)$ and $\tan\frac{\Delta}{2}/\big(\frac{2}{\sin\Delta} + \tan\frac{\Delta}{2}\big)$, respectively\;
            $gates \gets gates \cup \{(t,k,l)\}$\;
        }
        sort $gates$ in ascending order of the first entry
        \tcp*{time order the gates}
        $\hat{U}_{\rm Ad} \gets I$\;
        $s \gets 1$ 
        \tcp*{store sign of sampled TE-PAI channel}
        $t' \gets 0$\;
        \ForEach{$(t,k,l)\in gates$}{
            \eIf{$l=2$}{
                $\hat{U}_{\rm Ad} \gets \exp\left[-\im\, {\rm sgn}(a_k)\Delta P_k/2\right] \exp\left[-\im(t-t')H_{\rm HF}\right]\hat{U}_{\rm Ad}$\;
            }{
                $\hat{U}_{\rm Ad} \gets \exp\left[-\im \pi P_k/2\right] \exp\left[-\im(t-t')H_{\rm HF}\right]\hat{U}_{\rm Ad}$\;
                $s \gets -s$\;
            }
            $t' \gets t$\;
        }
        $\hat{U}_{\rm Ad} \gets \exp\left[-\im(T-t')H_{\rm HF}\right]\hat{U}_{\rm Ad}$\;
        $\hat{O}_j \gets s\,\bra{\rm HF}\hat{U}_{\rm Ad}^\dag O\hat{U}_{\rm Ad}\ket{\rm HF}$\;
    }
    
    $\Gamma \gets \exp\big[2Tz(1)\|a\|_1\tan \frac{\Delta}{2}\big]$
    \tcp*{TE-PAI normalization factor}
    \Output{$\frac{\Gamma}{N_{\rm s}}\sum_{j=1}^{N_{\rm s}} \hat{O}_j$}
\end{algorithm}
Here we made two changes. First, we randomly apply the gates corresponding to the terms in $H_{\rm int}$ on top of the evolution under $H_{\rm HF}$, which can be implemented without Trotter error. This reduces sampling overhead. Second, we incorporate the Hamiltonian's time dependence in the sampling of times at which the random gates are applied.

With the approximate ground state prepared with TE-PAI, we estimate the ground-state energy via the Hadamard test. The whole circuit is depicted as
\begin{align}
\label{circ:hadamard_app}
\begin{quantikz}
    \lstick{\ket{+}}
    &
    & \ctrl{1}
    & \meter{X/Y}
    \\
    \lstick{\ket{\rm HF}}
    & \gate{\hat{U}_{\rm Ad}}
    & \gate{\hat{V}_H}
    &
\end{quantikz} .
\end{align}
We approximate $V_H=\e^{\im s H}$ by TETRIS (Time-Evolution Through Random Independent Sampling), reviewed in the next section. See Alg.~\ref{alg:tetris}.
The measurements in $X$ and $Y$ bases in the circuit~\eqref{circ:hadamard_app} respectively calculate the real and imaginary parts of, 
\begin{align}
    \Gamma\, \Gamma_{\rm TET}\,\e^{-\im s E_{\rm HF}}\mathbb{E}\big[s_{\rm Ad}\,\Tr\big[\hat{V}_H \hat{\cal U}_{\rm Ad}[\ket{\rm HF}\bra{\rm HF}]\big]\big]
    =
    \e^{-\im s E_{\rm HF}}\bra{\rm HF}U_{\rm Ad}^\dag V_H U_{\rm Ad}\ket{\rm HF},
\end{align}
where $\mathbb{E}[\cdot]$ is the average over TE-PAI and TETRIS circuit instances,
$\Gamma=\e^{2Tz(1)\|a\|_1\tan \frac{\Delta}{2}}$ and $\Gamma_{\rm TET}=\e^{s\|a\|_1\tan \frac{\Delta}{2}}$ are normalization factors, and $s_{\rm Ad}$ is the sign associated with each TE-PAI channel.
This allows us to read off,
\begin{align}
    {\rm Im}\big[\e^{-\im s E_{\rm HF}}\bra{\rm HF}U_{\rm Ad}^\dag V_H U_{\rm Ad}\ket{\rm HF}\big]
    \approx
    {\rm Im}\big[\e^{\im s (E_{\rm exact} - E_{\rm HF})} \big]
    =
    \sin[s(E_{\rm exact} - E_{\rm HF})],
\end{align}
where $E_{\rm exact}$ is the exact ground state energy.

\section{Implementation of \texorpdfstring{$V_H$}{V\_H} with TETRIS}
\label{app:tetris}

TETRIS is a stochastic algorithm to implement a unitary operator $V_H=\e^{\im s H}$, where $H=\sum_k c_k P_k$ is a Hamiltonian decomposed into Pauli strings $P_k$ with coefficients $c_k\in\mathbb{R}$~\cite{Granet2023}.
As in the adiabatic evolution with TE-PAI, we split the Hamiltonian into $H=H_{\rm HF}+H_{\rm int}$~\eqref{eq:H_decomposition} and evolve under $H_{\rm int}$ stochastically.
According to Alg.~\ref{alg:tetris}, TETRIS constructs an operator $\hat{V}_H$ that approximates $V_H$ on average without bias via $\Gamma_{\rm TET}\bE[\hat{V}_H]=V_H$, at the cost of the sampling overhead, $\Gamma_{\rm TET}=\e^{s\|a\|_1\tan \frac{\Delta}{2}}$ with a free parameter $\Delta\in(0,\pi)$.
Similarly to TE-PAI, the larger the parameter $\Delta$ is, the lower the average gate count per circuit becomes, but the larger the sampling overhead.
It should be noted that TETRIS approximates a unitary operator instead of a quantum channel as TE-PAI does, making it suitable for the Hadamard test.

\begin{algorithm}
\caption{TETRIS}
\SetKwInOut{Input}{input}
\SetKwInOut{Output}{output}
\SetKwFor{ForEach}{for each}{do}{end}
\label{alg:tetris}
    \Input{
        Evolution time $s\in\mathbb{R}_{>0}$,
        interpolation angle $\Delta\in(0,\pi)$,
        number of samples $N_{\rm s}$,
        and Hamiltonian $H=H_{\rm HF}+H_{\rm int}$.
    }
    \For{$j=1$ \KwTo $N_{\rm s}$}{
        $gates \gets$ empty set\;
        $m\gets$ draw from Poisson distribution with parameter $s\|a\|_1/\sin\Delta$
        \tcp*{draw number of Pauli-rotation gates}
        \For{$i=1$ \KwTo $m$}{
            $t\gets$ draw uniformly from the interval $[0,s]$\;
            $k\gets$ draw from $\{1,\dots,K_{\rm int}\}$ with probability $|a_{k}|/\|a\|_1$\;
            $gates \gets gates \cup \{(t,k)\}$\;
        }
        sort $gates$ in ascending order of the first entry
        \tcp*{time order the gates}
        $\hat{V}_{H,j}\gets I$\;
        $t'\gets 0$\;
        \ForEach{$(t,k)\in gates$}{
            $\hat{V}_{H,j}\gets \exp\left[\im\,{\rm sgn}(a_k)\Delta P_k\right] \exp\left[\im(t-t')H_{\rm HF}\right] \hat{V}_{H,j}$\;
            $t'\gets t$\;
        }
        $\hat{V}_{H,j} \gets \exp\left[\im(s-t')H_{\rm HF}\right]\hat{V}_{H,j}$\;
    }
    $\Gamma_{\rm TET}\gets \exp\big[s\|a\|_1\tan \frac{\Delta}{2}\big]$
    \tcp*{TETRIS normalization factor}
    \Output{$\frac{\Gamma_{\rm TET}}{N_{\rm s}}\sum_{j=1}^{N_{\rm s}} \hat{V}_{H,j}$}
\end{algorithm}

\section{Additional numerical and experimental data for OTOCs of sparse SYK model}
\label{app:syk}

\begin{figure}
\includegraphics[width=0.33\linewidth]{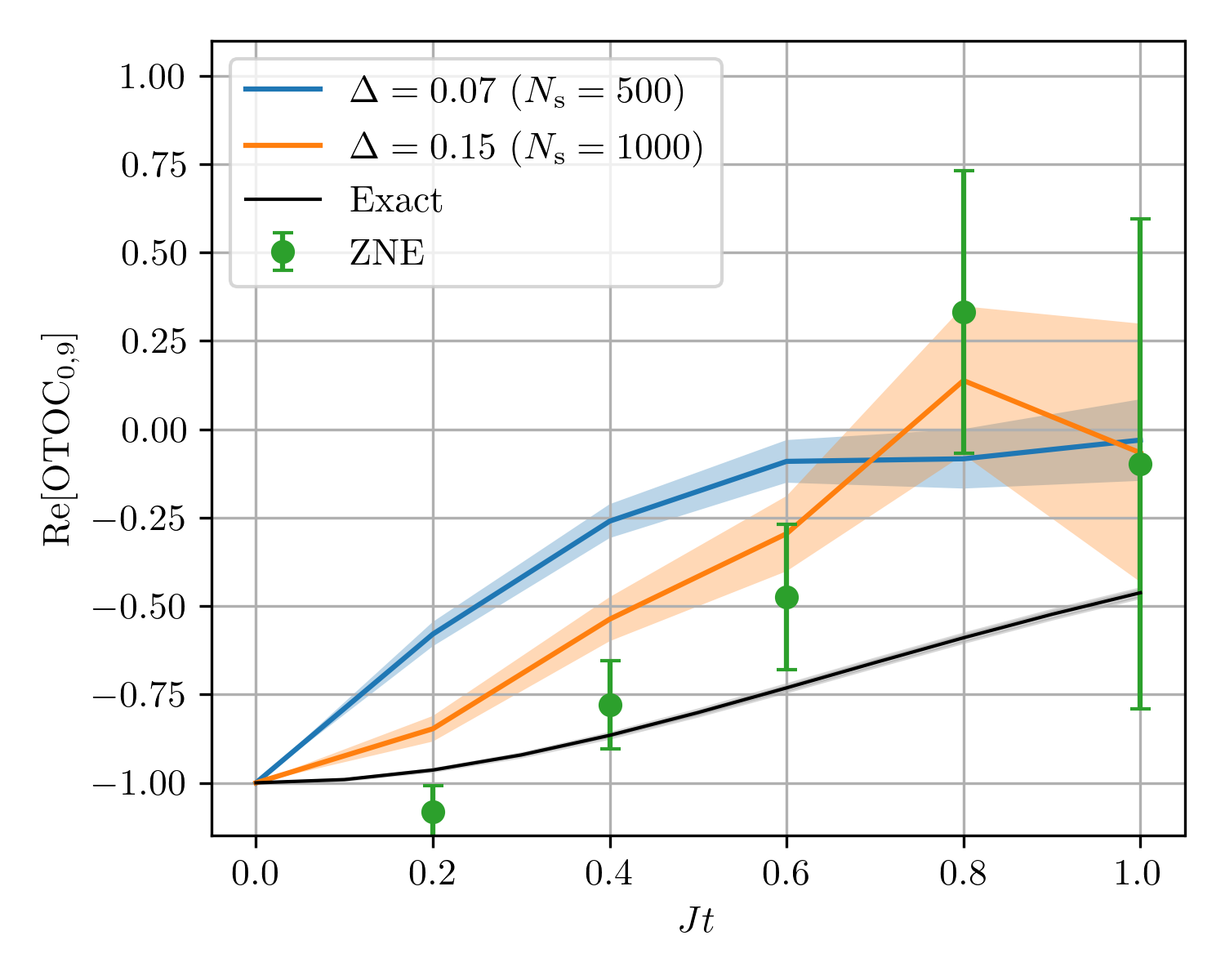}
\includegraphics[width=0.33\linewidth]{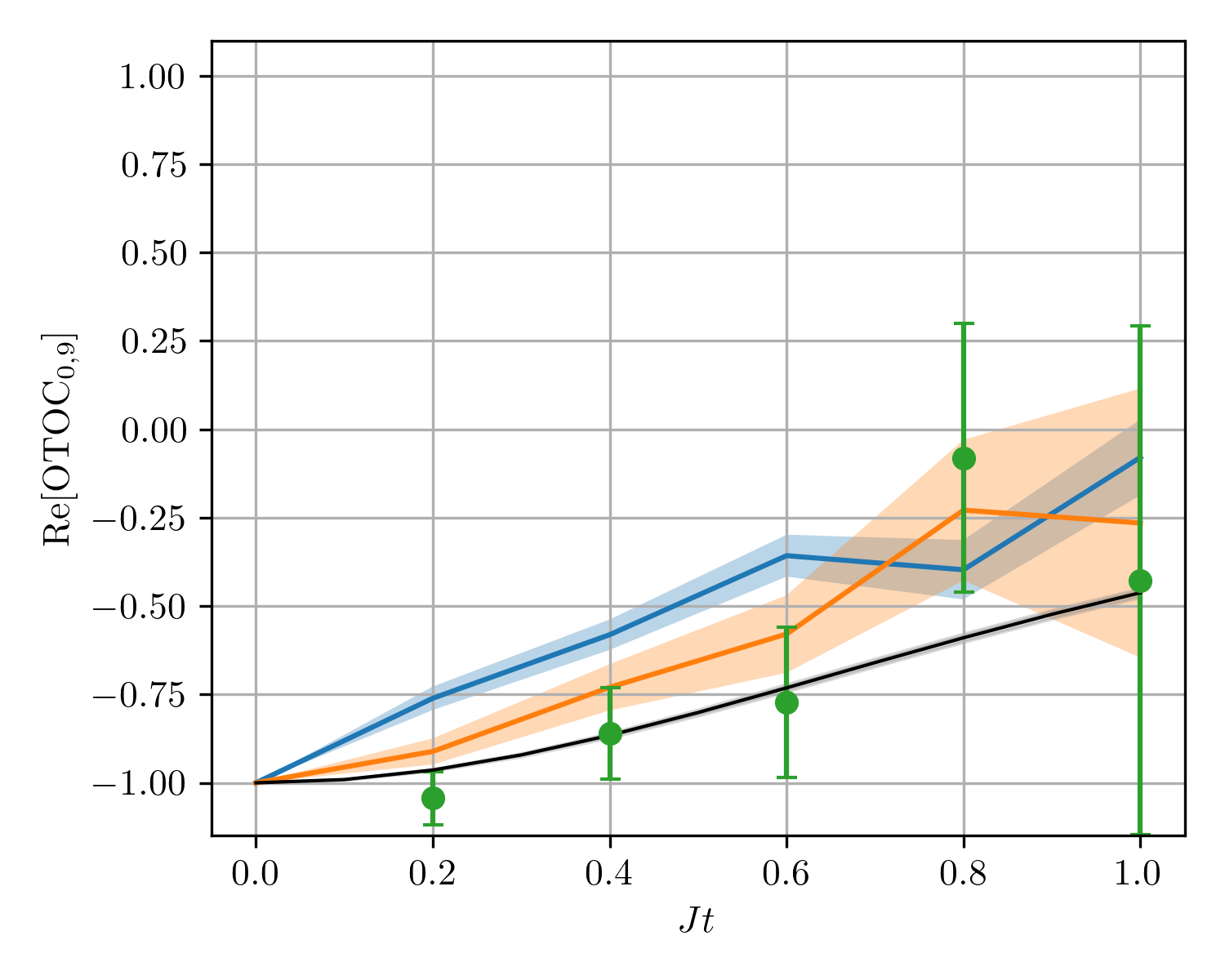}
\includegraphics[width=0.33\linewidth]{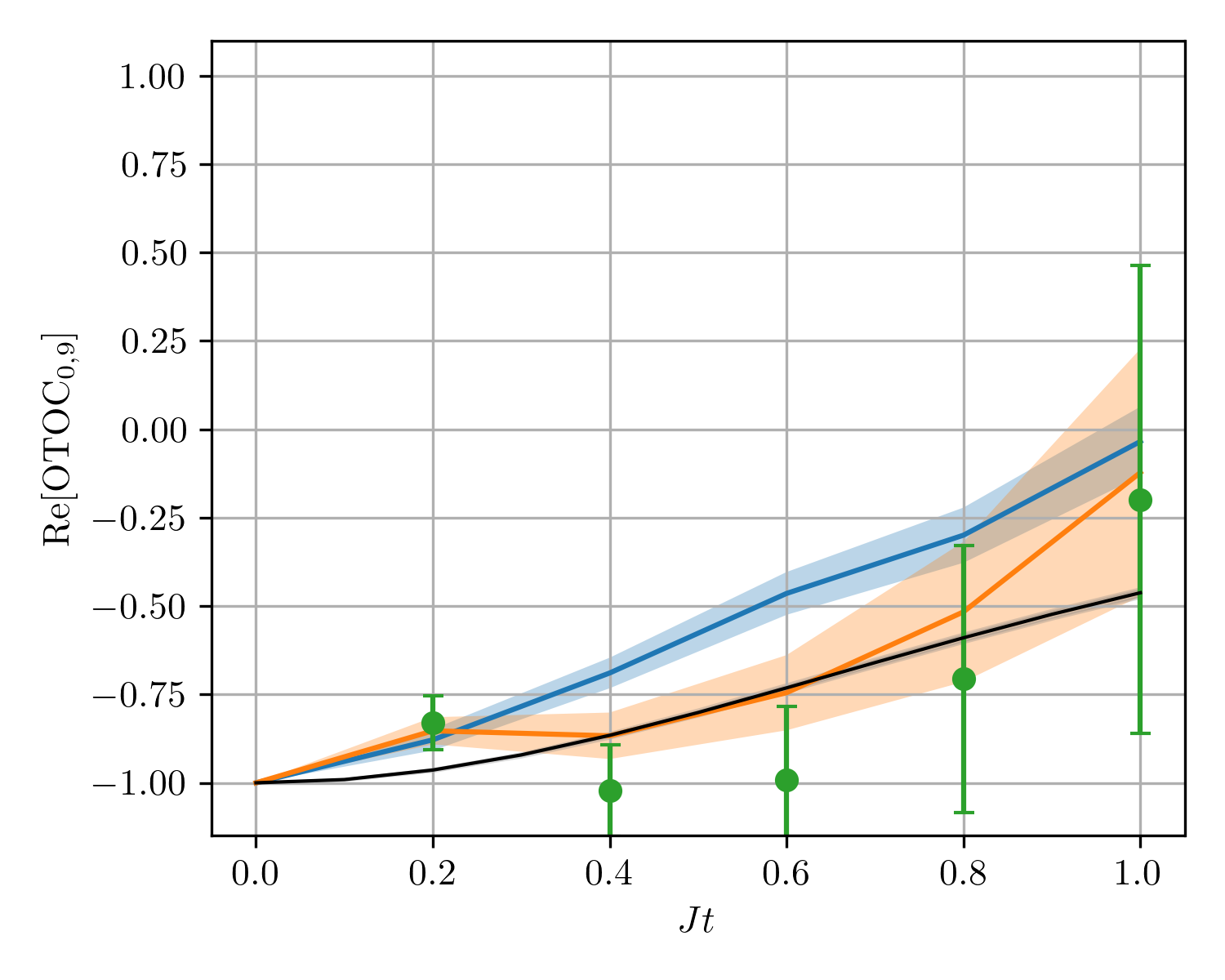}
\caption{\label{fig:SYK_ZNE_reimei}
    Results of OTOC calculation with ZNE on the Reimei quantum computer (left), its emulator without memory noise (middle), and simulator with single- and two-qubit-gate noise (right). The blue data are obtained by executing 500 circuits with $\Delta=0.07$, while the orange data are obtained by running 1000 circuits with $\Delta=0.15$. Each circuit is executed with 5 shots. The green data points represent the extrapolated values with ZNE. The OTOC computed by noiseless simulation of continuous dynamics is shown by the black solid curve, which is obtained by averaging over 500 random realizations of $J_{ijkl}$ and $p_{ijkl}$, with the gray region representing the associated statistical error.
}
\end{figure}

We provide the OTOC data for the sparse SYK model obtained by running the interferometric circuit in Fig.~\ref{circ:OTOC} in the main text on the Reimei quantum computer. See~\cite{Luo2017, Jafferis2022, Asaduzzaman2023, Granet2025} for simulations of the SYK model conducted on quantum computers.
The configuration of the SYK model is the same as that in the noisy simulation shown in Fig.~\ref{fig:SYK_ZNE_noisy_simulation} in the main text.
We post-select shots for which the parity of the measured bitstring on the system register is odd.
To facilitate further error mitigation with ZNE, we set the angle parameters in the randomized protocol to $\Delta=0.07$ and $\Delta=0.15$.
The results on Reimei are shown in the left panel of Fig.~\ref{fig:SYK_ZNE_reimei}. The agreement between experimental data and ideal data is poor.
This is likely due to the memory noise, that is, ions representing qubits during idling or transportation suffer from the hardware noise, modeled by the coherent noise channel, $\rho \mapsto \e^{-\im ftZ}\rho\e^{\im ftZ}$, where $t$ is the idling or transportation time and $f$ is a frequency parameter.
Since the SYK model contains long Pauli strings, the circuit depth is large and thus the idling time is long, leading to significant memory noise.
To confirm the effect of memory noise, we also run the same circuit on the Reimei emulator with memory noise turned off, and the result is shown in the middle panel of Fig.~\ref{fig:SYK_ZNE_reimei}.
The emulator data show a better agreement with the ideal data, confirming the effect of memory noise in the current experiment.
This suggests that the hardware results would be significantly improved if memory noise can be mitigated using techniques such as dynamical decoupling, which is not available on Reimei.
As in Fig.~\ref{fig:SYK_ZNE_noisy_simulation} in the main text, we also conduct the simulation only with the depolarizing noise with the strength $10^{-5}$ and $10^{-3}$ for each single- and two-qubit gate, respectively. The results are shown in the right panel of Fig.~\ref{fig:SYK_ZNE_reimei}. While the ZNE performs poorly, it is observed that the data for $\Delta=0.15$ mostly agrees with the exact data within statistical uncertainties.


\end{document}